\documentclass{amsart}

\usepackage{amsmath}
\usepackage{amscd}
\usepackage{amsfonts}
\usepackage{amssymb}
\usepackage{amsthm}
\usepackage{graphicx}
\usepackage{bbm}
\usepackage{ifpdf}
\usepackage{ifthen}
\usepackage{cite}

\newcommand{\ro}{\mathrm}

\newcommand{\R}{\mathbb{R}}
\newcommand{\C}{\mathbb{C}}
\newcommand{\N}{\mathbb{N}}
\newcommand{\Rn}[1][3]{\mathbb{R}^{#1}}
\newcommand{\Cn}[1][3]{\mathbb{C}^{#1}}
\newcommand{\Nn}[1][3]{\mathbb{N}^{#1}}

\newcommand{\bx}{\boldsymbol x}
\newcommand{\bgx}{\boldsymbol X^{\eps}}
\newcommand{\bgy}{\boldsymbol Y}
\newcommand{\by}{\boldsymbol y}
\newcommand{\bk}{\boldsymbol k}
\newcommand{\ba}{\boldsymbol a}
\newcommand{\be}{\boldsymbol \eta}
\newcommand{\bp}{\boldsymbol p}
\newcommand{\bv}{\boldsymbol v}
\newcommand{\bj}{\boldsymbol j}
\newcommand{\bgu}{\boldsymbol U}
\newcommand{\bxi}{\boldsymbol{\xi}}

\newcommand{\eps}{\varepsilon}
\newcommand{\sqeps}{\sqrt{\eps}}
\renewcommand{\phi}{\varphi}
\newcommand{\e}{\eta}

\renewcommand{\Im}{\ro{Im}}
\renewcommand{\Re}{\ro{Re}}
\renewcommand{\hat}{\widehat}
\renewcommand{\tilde}{\widetilde}

\newcommand{\nfourth}[1][3]{\frac{#1}{4}}
\newcommand{\nhalf}[1][1]{\frac{#1}{2}}
\newcommand{\skpr}[2]{\left\langle #1,\,#2\right\rangle}
\newcommand{\Dalpha}[1][\alpha]{\ro{D}^{#1}}

\newcommand{\Qp}[1][t]{\bgx(\bx_0,#1)}
\newcommand{\xa}[1][(t)]{\bx-\ba #1}

\newcommand{\vpsi}[1][]{\bv^{\psi^{#1}}}
\newcommand{\vpsk}[1][\bk]{\bv^{\psi^{\eps}_{#1}}}

\newcommand{\at}[1][(t)]{\ba #1}
\newcommand{\et}[1][(t)]{\be #1}
\newcommand{\Vat}[1][(t)]{V\left(\at[#1]\right)}
\newcommand{\Vtwo}{V^{(2)}}
\newcommand{\Vmxt}[2][(t)]{V_{#2}\left(\bx, \at[#1]\right)}

\newcommand{\psk}[1][\bk]{\psi^\eps_{#1}}
\newcommand{\phk}[1][\bk]{\phi_{#1}}
\newcommand{\Phk}[1][\bk]{\Phi^\eps_{#1}}

\newcommand{\PP}[1][\psk(\cdot,0)]{\mathbb{P}^{#1}}

\newcommand{\BA}[1][(t)]{B#1A#1^{-1}}

\newcommand{\At}[1][t]{A(#1)}
\newcommand{\Bt}[1][t]{B(#1)}

\newcommand{\Aadj}[1][j]{\mathcal{A}_{#1}^*(A,B,\eps,\ba,\be)}

\newcommand{\Heps}{H^{\eps}}
\newcommand{\Htildeps}[1][(t)]{\tilde{H}^{\eps} #1}
\newcommand{\Ueps}[1][t]{U^{\eps}(#1)}
\newcommand{\Utild}[1][t,s]{\tilde{U}^{\eps}(#1)}

\newcommand{\feps}[1][\cdot]{f^{\eps}_{m,\bk}(#1,s)}
\newcommand{\geps}[1][\cdot]{g^{\eps}_{m,\bk}(#1,t,s)}
\newcommand{\gepstild}[1][\cdot]{\tilde{g}^{\eps}_{m,\bk}(#1,t,s)}

\newcommand{\Gv}{G_V}
\newcommand{\Geps}[1][\bk,\delta]{G^{\eps}_{#1}}

\theoremstyle{plain}
\newtheorem*{thm}{Theorem}
\newtheorem{lem}{Lemma}
\newtheorem{cor}{Corollary}
\newtheorem{defi}{Definition}

\theoremstyle{definition}
\newtheorem{rem}{Remark}

\newenvironment{pro}[1][\negthickspace]{\noindent{\bf Proof #1.\:}}{\hfill $\Box$}

\newcommand{\Papersize}{letterpaper} 

\ifthenelse{\equal{\Papersize}{a4paper}}{\usepackage[dvips,includeall,papersize={210mm,
297mm},marginparsep=0cm,marginparwidth=0cm,left=2.5cm,textwidth=16cm,top=1.9cm,bottom=1.8cm]{geometry}}{}

\ifthenelse{\equal{\Papersize}{letterpaper}}{\usepackage[dvips,includeall,paper=letterpaper,marginparsep=0cm,marginparwidth=0cm,left=2.795cm,textwidth=16cm,top=2cm,bottom=2cm]{geometry}}{}

\begin{document}

\title[On the classical limit of Bohmian mechanics for Hagedorn wave packets]{\bf On the classical limit of Bohmian mechanics for Hagedorn wave packets}
\author[D.\ D\"urr]{Detlef D\"urr }
\author[S. R\"omer]{Sarah R\"omer}
\address{Mathematisches Institut der LMU,
Theresienstr. 39, 80333 M\"unchen, Germany}
\email{duerr@math.lmu.de, roemer@math.lmu.de}

\subjclass[2010]{81Q20, 46N50}
\keywords{Quantum mechanics, Bohmian mechanics, Classical limit, Semiclassical wave packets}

\begin{abstract}
We consider the classical limit of quantum mechanics in terms of Bohmian trajectories. For wave packets as defined by Hagedorn we show that the Bohmian trajectories converge to Newtonian trajectories in probability.

\end{abstract}
\maketitle

\section{Introduction}
\label{sec.intro}
There are many ways to formulate the classical limit of quantum mechanics. The strongest assertion would be about ``quantum particle trajectories'' becoming Newtonian. Particle trajectories, however, are not ontological elements of orthodox quantum theory and thus the ``classical limit'' must  be defined in some operational way. In contrast, Bohmian mechanics, which for all practical purposes is equivalent to quantum mechanics, is a quantum theory of point particles moving, so the study of the classical limit becomes a straightforward task \cite{allori:02,duerrteufel:09}: Under which circumstances are the Bohmian trajectories of particles approximately Newtonian trajectories? Here ``approximately'' can be understood in various manners. The technically simplest but also weakest is that at every time $t$ the Bohmian particle's position is close to the center of a ``classically moving''  very narrow  wave packet $\psi$. This essentially amounts to showing that $|\psi(t)|^2 $ is more or less transported along a Newtonian flow (see \cite{markowich:10} for a recent work on this).

The strongest and clearly most direct assertion would be that almost every Bohmian trajectory converges to a Newtonian trajectory in the uniform topology. We shall prove here a slightly weaker statement, namely  that the uniform closeness holds in probability. We shall establish this result for a particular class of wave packets which were defined by Hagedorn in \cite{hagedorn:98} and which move along classical paths.

To formulate the precise result let us recall that in Bohmian mechanics the state of a particle is described by a wave function $\psi(\by,s)$, where $\by\in\R^3,\,s\in\R$, and by its position $\bgy\in\R^3$. The wave function evolves according to Schr\"{o}dinger's equation ($\hbar=m=1$)
\begin{equation}
\label{eq.schroedinger}
i\frac{\partial}{\partial s}\psi(\by,s)=H\psi(\by,s):=\left(H_0+V(\by)\right)\psi(\by,s)
      :=\left(-\nhalf\triangle_y+V(\by)\right)\psi(\by,s)
\end{equation}
with the potential\footnote{More rigorously: $H$ is a self-adjoint extension of $H|_{C^\infty_{0}(\Omega)}=-\nhalf\triangle+V$ (with $V:\Omega\subseteq\R^3\rightarrow\R)$ on the Hilbert space $L^2(\R^3)$ with domain $\mathcal{D}(H)$.} $V$. The wave function governs the motion of the particle by
\begin{equation}
\label{eq.eq-of-motion}
\frac{d}{ds}\bgy(\by_0,s)=\vpsi(\bgy(\by_0,s),s)
      :=\Im\left(\frac{\nabla_y\psi(\bgy(\by_0,s),s)}{\psi(\bgy(\by_0,s),s)}\right)\,,
      \quad\bgy(\by_0,0)=\by_0\,.
\end{equation}
For a wave function $\psi$ the position $\bgy$ is a  random variable the distribution of which is given by the equivariant
probability measure $\PP[\psi]$ with density $|\psi(\by)|^2$ (Born's statistical
rule; see \cite{duerr:92, duerrteufel:09} for a precise assertion). This means that at any time $t$ the particle will typically be somewhere in the ``main'' support of $|\psi(\by,t)|$. Thus for a narrow wave packet which, according to Ehrenfest's theorem, moves -- at least for some time -- along a classical trajectory, at every instance of time $t$ the position of the particle will typically be close to a classical position. To be sure: this does not imply that a typical Bohmian trajectory stays close to the classical trajectory for the whole duration of a given time interval, since it may every now and then make a large excursion.

We shall consider a sufficiently smooth potential and a special class of initial wave functions where the potential $V$ varies on a much larger scale than the wave functions, see e.g. \cite{allori:02} for a physical discussions of the scales. More precisely, we choose $V^\eps(\by):= V(\eps\by)$ for some small parameter $\eps$, thus defining a microscopic $(\by,s)$ and a macroscopic scale $(\bx,t):=(\eps\by,\eps s)$. As initial wave functions we take the semiclassical wave packets $\Phk(\ba(0),\be(0),\cdot)$ defined by Hagedorn in \cite{hagedorn:85,hagedorn:98}. They are  non-isotropic three dimensional generalized Hermite polynomials of order $k:=|\bk|$ multiplied by a Gaussian wave packet centered around the classical phase space point $(\ba(0),\be(0))$. On the macroscopic scale, i.e.\ on the scale of variation of the potential, their standard deviation is of order $\sqeps$ both in position and momentum, that is they vary on an intermediate scale. This is the best order of $\eps$ allowed, since by Heisenberg's uncertainty relation $\sigma_{\by}\sigma_{\bp}\sim 1$ on the microscopic scale, so on the macroscopic scale $\sigma_{\bx}\sigma_{\bp}=\eps\sigma_{\by}\sigma_{\bp}$ must be of order $\eps$.

In the following, we change to macroscopic coordinates $(\bx,t)=(\eps\by,\eps s)$. With $\triangle:=\triangle_x$, $\nabla:=\nabla_x$ and $\psk[](\bx,t):=\eps^{-\nhalf[3]}\psi(\frac{\bx}{\eps},\frac{t}{\eps})$ Schr\"odinger's equation then reads
\begin{equation}
\label{eq.schr-gl-macro}
i\eps\frac{\partial}{\partial t}\psk[](\bx,t)=\Heps\psk[](\bx,t)=
        \left(-\frac{\eps^2}{2}\triangle+ V(\bx)\right)\psk[](\bx,t)\,.
\end{equation}
In this setting Hagedorn \cite{hagedorn:85,hagedorn:98} proved: With an error of order $\sqeps$ in $L^2$-norm the solution $\psk(\bx,t)$ of \eqref{eq.schr-gl-macro} with initial data $\psk(\bx,0)=\Phk(\ba(0),\be(0),\bx)$ is given by $\Phk(\at,\et,\bx)$, where $(\at,\et)$ is the corresponding classical phase space trajectory, that is the solution of the Newtonian law of motion with initial data $(\ba(0),\be(0))$.

Now consider the Bohmian trajectories on the macroscopic scale,i.e.\ solutions of the differential equation
\begin{equation}
\label{eq.eq-of-motion-macro}
\frac{d}{dt}\Qp=\vpsi(\Qp,t)
      =\eps\Im\left(\frac{\nabla\psk[](\Qp,t)}{\psk[](\Qp,t)}\right)\,,
      \quad\Qp[0]=\bx_0\,.
\end{equation}
Our main result is their convergence in probability: For all $T>0$ and $\gamma>0$ there exists some $R<\infty$ such that
\begin{equation*}
\PP(\{\bx_0\in\Rn\mid\max\limits_{t\in[0,T]}|\Qp-\at|\leq R\sqeps\})>1-\gamma
\end{equation*}
for all $\eps$ small enough.
\medskip

It is clearly desirable to have an analogous result for the velocities, so that convergence of ``phase space'' trajectories is achieved. However, the control of velocities introduces further technicalities. We shall  shortly discuss and present some results on the convergence of velocities in Section \ref{sec.velocities}. Next (Section \ref{sec.math-frame}) we give the mathematical setup: We briefly introduce the  dynamics  we want to compare and Hagedorn's result that we shall refine for our needs. Section \ref{sec.traj} describes our result on the classical limit. Proofs are in  Section \ref{sec.proof}.


\section{Mathematical framework}
\label{sec.math-frame}

\begin{defi}
\label{def.(V)^m} The potential  $V\in C^\infty(\Rn,\R)$ is in $\Gv$ if for all multi-indices $\alpha\in\Nn$
\begin{equation}
\label{eq.DV-bounded}
\max_{|\alpha|\leq 4} \|\Dalpha V\|_{\infty}\leq C_V
\end{equation}
for some $C_V <\infty$ and if multiplication by $V$ maps the Schwartz space $\mathcal{S}(\Rn)$ into itself, i.e.\ if $V\negthinspace f\in\mathcal{S}(\Rn)$ for all $f\in\mathcal{S}(\Rn)$. Here $\Dalpha$ denotes the (weak) derivative $\partial_{x_1}^{\alpha_1}\partial_{x_2}^{\alpha_2}\partial_{x_3}^{\alpha_3}$.
\end{defi}

The  requirement that $V$ maps $\mathcal{S}$ into itself is needed to get $\PP[\psi]$-almost sure global existence of Bohmian mechanics \cite{berndl:95,teufeltumulka:05} for initial wave functions $\psi\in\mathcal{S}$.

The quantum dynamics is given by Bohmian mechanics, i.e.\ by \eqref{eq.schr-gl-macro} and \eqref{eq.eq-of-motion-macro}. Equivariance of the measure $\PP[\psi]$ means that if $\bgx(\cdot,0)$ is $|\psi(\cdot,0)|^2$-distributed then $\bgx(\cdot,t)$ is $|\psi(\cdot,t)|^2$-distributed \cite{duerr:92}.
By $\Ueps$ we denote the unitary propagator generated by $\Heps$:
\begin{equation}
\label{eq.def_Ueps}
\frac{d}{dt}\Ueps|_{t=0}=-\frac{i}{\eps}\Heps\,.
\end{equation}

The classical dynamics is given by Newtonian mechanics, so the classical state of a particle at the macroscopic time $t$ is given by its classical position and velocity at that time, which we denote by $(\at, \et)$. For any given initial value $(\ba(0),\,\be(0))$ it is the unique global solution of Hamilton's equations of motion:
\begin{equation}
\label{eq.clmotion}
\begin{split}
&\dot\ba(t)=\et\,,\\
&\dot\be(t)=-\nabla \Vat\,.
\end{split}
\end{equation}

We introduce now the class of  Hagedorn's wave functions for which we shall establish the classicality of Bohmian trajectories. Hagedorn's wave packets are the eigenfunctions of some generalized 3-dimensional harmonic oscillator. As such they are generalized Hermite functions, i.e.\ products of generalized Hermite polynomials and the Gaussian ground state $\phi_0$. More precisely, for every $\eps>0$, every phase space point $(a,\e)$ and every pair of admissible matrices $(A,B)\in\C^{3\times 3}$ Hagedorn constructed an orthonormal basis of $L^2(\Rn)$ consisting of semiclassical wave packets
\begin{gather}
\label{eq.def-phk}
\phk(A,B,\eps,\ba,\be,\bx):=\frac{1}{\sqrt{\bk!}}\Aadj[]^{\bk}\phk[0](A,B,\eps,\ba,\be,\bx)\,,\qquad\bk\in\Nn,\\
\intertext{with the ground state}
\label{eq.def-ph0}
\phk[0](A,B,\eps,\ba,\be,\bx):=
(\pi\eps)^{-\nfourth}\det(A)^{-\nhalf[1]}
      \exp\left[-\frac{1}{2\eps}\skpr{(\xa[])}{\BA[](\xa[])}+\frac{i}{\eps}\skpr{\et[]}{(\xa[])}\right]\\
\intertext{and the formal vector of raising operators}
\notag
\Aadj[]:=\frac{1}{\sqrt{2\eps}}\left[B^*(\xa[])-iA^*(\bp-\be)\right]\,.
\end{gather}
Here $\bp=-i\nabla_y=-i\eps\nabla$, $\skpr{\cdot}{\cdot}$ is the canonical scalar product on $\Cn[n]$ and $(A,B)$ are admissible if
\begin{equation}
\label{eq.cond_AB}
A^TB-B^TA=0
\quad\text{and}\quad
A^*B+B^*A=2\,.
\end{equation}
In particular, \eqref{eq.cond_AB} implies that $A$ is invertible, $\Re(\BA[])=\left(AA^*\right)^{-1}$ and thus that for some constants $0<C\leq\tilde{C}<\infty$
\[
\eps^{-\nfourth}Ce^{-C\frac{\|\bx-\ba\|^2}{\eps}}\leq|\phk[0](A,B,\eps,\ba,\be,\bx)|
      \leq\eps^{-\nfourth}\tilde{C}e^{-\tilde{C}\frac{\|\bx-\ba\|^2}{\eps}}\,.
\]

Moreover, for any multi-index $\alpha\in\Nn$
\begin{gather}
\label{eq.expansion-(x-a/sqeps)Phk}
\left(\frac{\xa[]}{\sqeps}\right)^\alpha\phk(A,B,\eps,\ba,\be,\bx)
      =\hspace{-0.7 cm}\sum_{\substack{|\bk-\bk'|\leq|\alpha|\\|\bk-\bk'|+|\alpha|\text{ even}}}\hspace{-0.6 cm}
            C^\alpha_{\bk\bk'}(A)\phk[\bk'](A,B,\eps,\ba,\be,\bx)\\
\intertext{and}
\label{eq.expansion-(p-e/sqeps)Phk}
\left(\frac{\bp-\be}{\sqeps}\right)^\alpha\phk(A,B,\eps,\ba,\be,\bx)
      =\hspace{-0.7 cm}\sum_{\substack{|\bk-\bk'|\leq|\alpha|\\|\bk-\bk'|+|\alpha|\text{ even}}}\hspace{-0.6 cm}
            C^\alpha_{\bk\bk'}(B)\phk[\bk'](A,B,\eps,\ba,\be,\bx)
\end{gather}
where $C^\alpha_{\bk\bk'}$ depends continuously on $A$ resp. $B$. This in turn entails, denoting by $\skpr{\cdot}{\cdot}$ also the scalar product on $L^2(\Rn)$,
\begin{gather}
\label{eq.expt_value_Phk}
\begin{split}
&\skpr{\phk(A,B,\eps,\ba,\be,\bx)}{\bx\,\phk(A,B,\eps,\ba,\be,\bx)}=\at\,,\\
&\skpr{\phk(A,B,\eps,\ba,\be,\bx)}{\bp\,\phk(A,B,\eps,\ba,\be,\bx)}=\et
\end{split}\\
\intertext{and, for any multi-index $\alpha\in\Nn$,}
\label{eq.higher_mom_Phk}
\begin{split}
&\left\|(\xa[])^\alpha\phk(A,B,\eps,\ba,\be,\bx)\right\|_2\leq C^\alpha_{\bk}(A)\eps^{\nhalf[|\alpha|]}\,,\\
&\left\|(\bp-\be)^\alpha\phk(A,B,\eps,\ba,\be,\bx)\right\|_2\leq C^\alpha_{\bk}(B)\eps^{\nhalf[|\alpha|]}\,.
\end{split}
\end{gather}

The $\phk$s and their gradients scale in $\eps$ as follows: There is a constant $C<\infty$, depending on $\bk,\,A$ and $B$ such that
\begin{gather}
\label{eq.scale_Phk}
\begin{split}
\left|\phk(A,B,\eps,\ba,\be,\bx)\right|=\eps^{-\nfourth}\Big|\phk\Big(A,B,&1,0,0,\frac{\xa}{\sqeps}\Big)\Big|\\
      &\leq\eps^{-\nfourth}
            C\Big(1+\frac{|\xa|}{\sqeps}\Big)^{k}e^{-\nhalf C\left(\frac{|\xa|}{\sqeps}\right)^2}\\
\end{split}
\intertext{and}
\label{eq.scale_nabla_Phk}
\begin{split}
\Big|\Big(\nabla-\frac{i}{\eps}\et\Big)\phk(A,B,\eps,\ba,\be,\bx)\Big|
            =\eps^{-\nhalf}\Big|&\Big(\frac{\bp-\be}{\sqeps}\Big)\phk(A,B,\eps,\ba,\be,\bx)\Big|\leq\\
    &\leq\eps^{-\nfourth[5]}C\Big(1+\frac{|\xa|}{\sqeps}\Big)^{k+1}
            e^{-\nhalf C\left(\frac{|\xa|}{\sqeps}\right)^2}
\end{split}
\end{gather}
for all $\bx\in\Rn$.

Hagedorn's wave packets yield approximate solutions to Schr\"odinger's equation \eqref{eq.schr-gl-macro}: Let $\at,$ $\et$ be a solution of \eqref{eq.clmotion} and $\At,\,\Bt$ a solution of
\begin{equation*}
\begin{aligned}
\dot{A}(t)&=i\Bt\,,\\
\dot{B}(t)&=i\Vtwo(\at)\At\,,
\end{aligned}
\end{equation*}
with initial data $A(0),\,B(0)$ fulfilling \eqref{eq.cond_AB}. Call
\begin{gather}
\label{eq.def-Phk}
\Phk(\bx,t):=e^{\frac{i}{\eps}S(t)}\phi_{\bk}(\At,\Bt,\eps,\at,\et,\bx)\\
\intertext{the semiclassically time evolved wave packet and}
\label{eq.def-psk}
\psk(\bx,t):=\Ueps\Phk(\bx,0)=\Ueps\phi_{\bk}(A(0),B(0),\eps,a(0),\e(0),\bx)
\end{gather}
the Schr\"odinger evolved wave packet, where $S(t)=\int\limits_{0}^t\left[\nhalf\e^2(s)-V(\at[(s)])\right]\negthinspace ds$ is the usual classical action. Then for every $T>0$ there is some $C<\infty$ (depending on $T,\bk,\at,\et,\At$) such that
\begin{equation}
\label{eq.psk-Phk-L2}
\|\psk(\bx,t)-\Phk(\bx,t)\|_2<C\sqeps
\end{equation}
for all $t\in[0,T]$ (\cite{hagedorn:98} Theorem 3.5).

Moreover, the semiclassical evolution  of the  packet is of Schr\"odinger type: Define the truncated, time dependent quadratic Hamiltonian
\begin{gather}
\label{eq.def_Htild}
\Htildeps:=\Htildeps[(\at)]:=-\nhalf[\eps^2]\triangle + \Vmxt{0,2}\\
\intertext{with $V_{0,2}$ the quadratic approximation of $V$ at $\at$,}
\notag
\Vmxt{0,2}:=\sum_{|\alpha|=0}^{2}\frac{1}{\alpha!}\left(\Dalpha V\right)(\at)(\bx-\at)^{\alpha},
\end{gather}
and let $\Utild$ the unitary propagator generated by $\Htildeps[]$, i.e. with
\begin{equation}
\label{eq.def_Utild}
\frac{d}{dt}\Utild|_{t=s}=-\frac{i}{\eps}\Htildeps[(s)]\;.
\end{equation}
Then
\begin{equation}
\label{eq.time_ev_Phk}
\Phk(\bx,t)=\Utild\Phk(\bx,s)
\end{equation}
for any $t\,,s\in\R$ (\cite{hagedorn:98} Theorem 3.4).

\section{Bohmian trajectories of Hagedorn wave packets}
\label{sec.traj}
\begin{thm}
Let $\Heps=-\nhalf[\eps^2]\triangle + V(\bx)\,,\;\mathcal{D}(\Heps)\subset L^2(\Rn)$ with $V\in \Gv$. For $\bk\in\Nn$ let $\psk(\bx,t)$ be given by \eqref{eq.def-psk}, $\bgx(\bx_0,t)$ by \eqref{eq.eq-of-motion-macro} and $\at$ by \eqref{eq.clmotion}. Then
\begin{enumerate}
\item
For all $\eps>0$ the Bohmian trajectories $\Qp$ exist globally in time for $\PP$-almost all initial positions $\bx_0\in\Rn$.
\item
For all $T>0$ and all $\gamma>0$ there exists some $R<\infty$ and some $\eps_0>0$ such that
\begin{equation}
\label{eq.P(Q_good)>1-gamma}
\PP(\{\bx_0\in\Rn\mid\,\max\limits_{t\in[0,T]}|\Qp-\at|\leq R\sqeps\})>1-\gamma
\end{equation}
for all $0<\eps\leq\eps_0$.
\end{enumerate}
\end{thm}
\noindent For the proof we shall use that the probability that a Bohmian trajectory crosses a certain surface (here the moving sphere $S_{R\sqeps}(\at)$) is bounded by the quantum probability flux  $\bj^{\psk}=\vpsk|\psk|^2=\Im[(\psk)^*\nabla\psk]$ across this surface (Subsection \ref{subsec.proof_theorem}). For that we  need pointwise estimates on the quantum probability current density, i.e. on $\psk$ and $\nabla\psk$.
\begin{lem}
\label{lem.pointw_est_psk}
Let $\Heps=-\nhalf[\eps^2]\triangle + V(\bx)\,,\;\mathcal{D}(\Heps)\subset L^2(\Rn)$ with $V\in \Gv$. For $\bk\in\Nn$ let $\Phk(\bx,t)$ be given by \eqref{eq.def-Phk} and $\psk(\bx,t)$ by \eqref{eq.def-psk}. Then for all $T>0$ there exists some $C<\infty$ such that
\begin{gather}
\label{eq.ptws_est_psk}
\max_{t\in[0\,,T]}\left\|\psk(\cdot\,,t)-\Phk(\cdot\,,t)\right\|_{\infty}\leq C\eps^{-\nfourth[1]}\\
\intertext{and}
\label{eq.ptws_est_nablapsk}
\max_{t\in[0\,,T]}\left\|\,\left|\nabla\psk(\cdot\,,t)-
        \nabla\Phk(\cdot\,,t)\right|\,\right\|_{\infty}\leq C\eps^{-\nfourth[5]}
\end{gather}
where $\|\cdot\|_{\infty}=\sup\limits_{\bx\in\Rn}|\cdot|$.
\end{lem}
For the proof see Subsection \ref{subsec.proof-lem-ptws-est}. Note that, since $\|\Phk\|_\infty\stackrel{\eqref{eq.scale_Phk}}{\sim}\eps^{-\nfourth}$ resp.\ $\|\nabla\Phk\|_\infty\stackrel{\eqref{eq.scale_nabla_Phk}}{\sim}\frac{1}{\eps}\|\Phk\|_\infty+\eps^{-\nfourth[5]}\sim\eps^{-\nfourth[7]}$, the relative value of the differences $\|\psk-\Phk\|_\infty$ resp.\ $\|\nabla(\psk-\Phk)\|_\infty$ is of order $\sqeps$ each.


\section{What about velocities?}
\label{sec.velocities}

The theorem above is a result about a particle's typical Bohmian position as a function of time. To extend this to velocities, i.e.\ to show that also
\begin{equation}
\label{eq.P(v-good)>1-gamma}
\PP(\{\bx_0\in\Rn\mid\,\max\limits_{t\in[0,T]}|\vpsk\left(\Qp,t\right)-\et|\leq K\sqeps\})>1-\gamma
\end{equation}
for some $K<\infty$ and all $\eps$ small enough, one needs to control the probability that the Bohmian trajectory comes too close to the wave function's nodes where the velocity field $\vpsk=\eps\Im\left(\frac{\nabla\psk}{\psk}\right)$ is ill defined. More precisely, since by \eqref{eq.scale_nabla_Phk} and Lemma \ref{lem.pointw_est_psk}
\begin{equation*}
\big|\vpsk-\be\big|=\Big|\Im\frac{\eps\nabla\psk-i\be\psk}{\psk}\Big|
      \leq\frac{|\eps\nabla\Phk-i\be\Phk|+\eps|\nabla\psk-\nabla\Phk|+\e|\psk-\Phk|}{|\psk|}
      \sim\frac{\sqeps}{\eps^{\nfourth}|\psk|}\,,
\end{equation*}
one needs that there exists some $\delta_{T,\bk}(\gamma)>0$ such that
\begin{equation}
\label{eq.P(Q-uniformally-far-away-from-nodes)>1-gamma}
\PP\left(\left\{\bx_0\in\Rn\mid\;\big|\psk\big(\Qp,t\big)\big|>\eps^{-\nfourth}\delta_{T,\bk}(\gamma)
      \text{ for all }t\in[0,T]\right\}\right)>1-\gamma
\end{equation}
\emph{for all} $\eps$ small enough. From the $\PP$-almost sure global existence of Bohmian mechanics \cite{berndl:94,teufeltumulka:05} one has that for all $\eps>0$ there is some $\delta^\eps_{\bk}(\gamma)>0$ such that
\begin{equation}
\label{eq.P(Q-away-from-nodes)>1-gamma}
\PP\left(\left\{\bx_0\in\Rn\mid\;|\psk(\Qp,t)|>\delta^\eps_{\bk}(\gamma)
      \text{ for all }t\in\R\right\}\right)>1-\gamma\,.
\end{equation}
However, we need more, namely the $\eps$-dependence of $\delta^\eps_{\bk}(\gamma)$. This may be achieved by  scrutinizing the existence proof, in particular the proof  of \eqref{eq.P(Q-away-from-nodes)>1-gamma} in \cite{berndl:94}. We shall not do so here. Instead, we note that for the ground state $\bk=0$ \eqref{eq.P(v-good)>1-gamma} is an easy corollary of our theorem and Lemma \ref{lem.pointw_est_psk}. This is due to the fact that $\Phk[0]$ is just a Gaussian and thus does not possess any nodes.  Similarly, also WKB--wave functions do not possess nodes. See \cite{markowich:10} for an assertion  concerning on Bohmian velocities in that case.

For the Gaussian ground state
\begin{gather*}
|\psk[0](\bx,t)|\stackrel{\eqref{eq.ptws_est_psk}}{\geq}|\Phk[0](\bx,t)|-C\eps^{-\nfourth[1]}
      \stackrel{\eqref{eq.def-ph0}}{\geq}C\eps^{-\nfourth}(e^{-CR}-\sqeps)\geq C\eps^{-\nfourth}\\
\intertext {and thus}
\big|\vpsk[0](\bx,t)-\et\big|\leq C\sqeps
\end{gather*}
whenever $|\bx-\at|\leq\sqeps R$ and $\eps$ small enough. So our theorem gives
\begin{cor}
\label{cor.traj+vel_for_k=0}
Under the same assumptions as in the theorem for all $T>0$ and all $\gamma>0$ there exist some $R<\infty,\;K<\infty$ and some $\eps_0>0$ such that
\begin{equation}
\label{eq.P(Q_good,v_good)>1-gamma}
\begin{split}
\PP[{\psk[0]}(\cdot,0)]\Big(\big\{\bx_0\in\Rn\mid\,\max\limits_{t\in[0,T]}|\Qp-&\at|\leq R\sqeps\\
        &\wedge\;\max\limits_{t\in[0,T]}|\vpsk[0](\Qp,t)-\et|\leq K\sqeps\big\}\Big)>1-\gamma
\end{split}
\end{equation}
for all $0<\eps\leq\eps_0$.
\end{cor}
A weaker statement which is true for any $\psk$ is the following. Since a typical Bohmian trajectory may not deviate too much from its corresponding classical one, the time averaged values of the velocities must be close: For any macroscopic time interval  $0<\delta t\leq\nhalf[T]$ define the time-averaged Bohmian and classical velocities ($t\in[\delta t,T-\delta t]$)
\begin{align*}
\vpsk_{\delta t}(\bx_0,t)&:=\frac{1}{2\delta t}\int\limits_{t-\delta t}^{t+\delta t}\vpsk\left(\Qp[s],s\right)ds\,,\\
\be_{\delta t}(t)&:=\frac{1}{2\delta t}\int\limits_{t-\delta t}^{t+\delta t}\et[(s)]ds\,.
\end{align*}
Now suppose $\bx_0\in\Rn$ is such that $\max\limits_{t\in[0,T]}|\Qp-\at|\leq R\sqeps$. Then
\begin{align*}
|\vpsk_{\delta t}(\bx_0,t)-\be_{\delta t}(t)|
            &=\frac{1}{2\delta t}\Big|\int\limits_{t-\delta t}^{t+\delta t}\hspace{-0.2cm}
            \big(\vpsk\left(\Qp[s],s\right)-\et[(s)]\big)ds\Big|\\
      &\leq\frac{1}{2\delta t}\big[\,|\Qp[t+\delta t]-\at[(t+\delta t)]|+|\Qp[t-\delta t]-\at[(t-\delta t)]|\,\big]\\
      &\leq\frac{R}{\delta t}\sqeps\,.
\end{align*}
So our theorem gives
\begin{cor}
\label{cor.P(v-averaged-good)>1-gamma}
Under the same assumptions as in the theorem for all $T>0,\,\gamma>0$ there exists some $R<\infty$ and some $\eps_0>0$ such that for any $0<\delta t\leq\nhalf[T]$
\begin{equation}
\label{eq.P(v-averaged-good)>1-gamma}
\PP\Big(\Big\{\bx_0\in\Rn\mid\,\max\limits_{t\in[\delta t,T-\delta t]}|\vpsk_{\delta t}(\bx_0,t)-\be_{\delta t}(t)|
      \leq \frac{R}{\delta t}\sqeps\,\Big\}\Big)>1-\gamma
\end{equation}
for all $0<\eps\leq\eps_0$.
\end{cor}

We conclude with a note on 
the Hamilton-Jacobi form of Bohmian mechanics. Setting
\[
\psi^\eps(\bx,t)=R^\eps(\bx,t)e^{\frac{i}{\eps}S^\eps(\bx,t)}\,,
\]
the real part of Schr\"odinger's equation \eqref{eq.schr-gl-macro} gives
\begin{gather*}
\partial_t S^\eps(\bx,t)+V(\bx)-\nhalf[\eps^2]\frac{\triangle R^\eps(\bx,t)}{R^\eps(\bx,t)}
      +\nhalf\left(\nabla S^\eps(\bx,t)\right)^2=0\\
\intertext{while \eqref{eq.eq-of-motion-macro} reads}
\frac{d}{dt}\Qp=\nabla S^\eps\left(\Qp,t\right)\,.
\end{gather*}
Except for the additional ``quantum potential'' $V^\eps_Q:=-\nhalf[\eps^2]\frac{\triangle R^\eps}{R^\eps}$ these are the classical Hamilton-Jacobi equations. This suggests very directly that Bohmian particles behave classically
whenever $V^\eps_Q$ is negligible \cite{bohm:52, holland:95, markowich:10}. However, due to the occurrence of $\frac{1}{R^\eps}$  a proof along these lines must deal with the nodes problem we discussed above.


\section{Proof}
\label{sec.proof}

\subsection{Proof of the theorem}
\label{subsec.proof_theorem}
$(i)$ is a direct consequence of Corollary 3.2 in \cite{berndl:95} resp.\ of Corollary 4 in \cite{teufeltumulka:05} if we can show that the initial wave function $\psk(\cdot,0)=\Phk(\cdot,0)$ is a $C^\infty$-vector of $\Heps$, $\Phk(\cdot,0)\in C^\infty(\Heps)=\bigcap_{n=1}^{\infty}\mathcal{D}\left((\Heps)^n\right)$. This is the case, since $\Phk(\cdot,0)\in\mathcal{S}(\Rn)$ and $V\in\Gv$ guarantees that $\Heps$ maps the Schwartz space $\mathcal{S}(\Rn)$ into itself and thus that $\mathcal{S}(\Rn)\subset C^\infty(\Heps)$.

$(ii)$: Let $\gamma>0$. For $\eps>0$ and $R<\infty$ define
\begin{equation*}
\Geps[R]:=\Big\{\bx_0\in\Rn\mid\,\max_{t\in[0,\,T]}|\Qp-\at|<R\sqeps\Big\}\,.
\end{equation*}
Our task is to show that, for suitable $R$ and $\eps$, $\PP\left((\Geps[R])^c\right)<\gamma$.

For this we split off the probability that a trajectory already starts too far off from the classical one:
\begin{equation}
\label{eq.PP(bad_traj)_geq_PP(bad_beginning)+PP(leaves_ball)}
\begin{split}
\PP&\left((\Geps[R])^c\right)\\
        &\hspace{-0.8cm}\leq\PP\left(\left\{\bx_0\in\Rn\mid\,|\bx_0-\ba(0)|\geq R\sqeps\right\}\right)\\
        &\hspace{-0.2cm}+\PP\left(\left\{\bx_0\in\Rn\mid\,|\bx_0-\ba(0)|< R\sqeps\:\wedge\:
                \exists\,t\in(0,T]:\,|\Qp-\at|\geq R\sqeps\right\}\right)\\
        &\hspace{-0.8cm}=:\PP\left(B_{R\sqeps}(\ba(0))^c\right)+\PP\left(\mathcal{M}^T_{R\sqeps}\right)\,.
\end{split}
\end{equation}
Then
\begin{equation*}
\begin{split}
\PP\left(B_{R\sqeps}(\ba(0))^c\right)&
            \,=\hspace{-0.6cm}\int\limits_{|\bx-\ba(0)|\geq R\sqeps}\hspace{-0.6cm}|\psk(\bx,0)|^2\,d^3x
            \,\stackrel{\eqref{eq.def-psk}}{=}\hspace{-0.6cm}\int\limits_{|\bx-\ba(0)|\geq R\sqeps}\hspace{-0.6cm}
                  |\phk(A(0),B(0),\eps,\ba(0),\be(0),\bx)|^2\,d^3x\\
      &\,\stackrel{\eqref{eq.scale_Phk}}{=}\hspace{-0.15cm}\int\limits_{y\geq R}\hspace{-0.15cm}
            |\phk(A(0),B(0),1,0,0,\by)|^2\,d^3y\,,
\end{split}
\end{equation*}
where in the last step we substituted $\by=\frac{\bx-a(0)}{\sqeps}$. Since $\phk(\At,\Bt,1,0,0,\cdot)$ is square summable (in fact it is normalized) there is some $R'>0$ \emph{independent} of $\eps$ such that
\begin{equation}
\label{eq.PP(bad_beginning)_leq_gammahalf}
\PP\left(B_{R\sqeps}(\ba(0))^c\right)\leq\nhalf[\gamma]
\end{equation}
for all $R>R'$.

Since $\Qp$ (as a solution of \eqref{eq.eq-of-motion}) is continuous in $t$, $x_0\in\mathcal{M}^T_{R\sqeps}$ implies that $\Qp[t]$ crosses the moving sphere $S_{R\sqeps}(\at)$ at least once and outwards in $(0,T]$. Therefore $\PP\big(\mathcal{M}^T_{R\sqeps}\big)$ is bounded from above by the probability that some
trajectory crosses $S_{R\sqeps}(\at)$ in any direction in $(0,T]$. In Subsection 2.3.2 of \cite{berndl:94}
Berndl invoked the probabilistic meaning of the quantum probability current density $J^\psi:=(\bj^{\psi},\,|\psi|^2)$ with $\bj^{\psi}:=\eps\Im\big(\psi^*\nabla\psi\big)$ to prove that the expected number of crossings\footnote{This also includes tangential "crossings" in which the trajectory remains on the same side of $\Sigma$.} through a smooth surface $\Sigma$ in configuration-space-time by the random configuration-space-time trajectory $\left(\bgx(\cdot,t)\,,\,t\right)$ is given by the modulus of the flux across this surface,
\begin{equation*}
\int\limits_{\Sigma}\left|J^\psi(\bx,t)\cdot \bgu\right|\,d\sigma\,,
\end{equation*}
where $\bgu$ denotes the local unit normal vector at $(\bx,t)$ (see also the argument given in \cite{berndl:95}, p. 11.). Since any trajectory $\left(\Qp\,,\,t\right)$ will cross $\Sigma$ an integral number of times (including $0$
and $\infty$) this expected value gives an upper bound for the probability that $\left(\Qp\,,\,t\right)$
crosses $\Sigma$. So in our case we obtain
\begin{equation}
\label{eq.PP(leaves_ball)_leq_FAS}
\PP\Big(\mathcal{M}^T_{R\sqeps}\Big)\leq\int\limits_{\Sigma^\eps_T}\Big|J^{\psk}(\bx,t)\cdot\bgu\Big|\,d\sigma
\end{equation}
where
\begin{equation*}
\Sigma^\eps_T=\{(\bx,\,t)\mid\,t\in[0,T],\,\bx\in S_{R\sqeps}(\at)\}
\end{equation*}
and, using spatial polar coordinates centered at $\at$, $\bgu=\frac{1}{\sqrt{1+\skpr{\et}{\hat{e}_r}^2}}\Big(\hat{e}_r,-\skpr{\et}{\hat{e}_r}\Big)$ and
$d\sigma=\sqrt{1+\skpr{\et}{\hat{e}_r}^2}\,\eps R^2d\Omega\,dt$. Here $\hat{e}_r=\left(\cos\phi\sin\theta,\sin\phi\sin\theta,\cos\theta\right)$ and $d\Omega=\sin\theta\,d\phi\,d\theta$. Thus
\begin{equation}
\label{eq.|J_U|dsigma}
\begin{split}
|J^{\psk}(\bx,t)\cdot \bgu|\,d\sigma&=
            \left|\skpr{\bj^{\psk}(\bx,t)-|\psk(\bx,t)|^2\et}{\hat{e}_r}\right|\eps R^2d\Omega\\
      &\hspace{4cm}\leq|\bj^{\psk}(\bx,t)-|\psk(\bx,t)|^2\et|\eps R^2d\Omega
\end{split}
\end{equation}
where $\bj^{\psk}(\bx,t)-|\psk(\bx,t)|^2\et$ is evaluated at points $(\bx,t)\in\Sigma^\eps_T$. By the definition of $\bj^{\psi}$ and since $\et$ is always real
\begin{equation*}
\begin{split}
\big|\bj^{\psk}-|\psk|^2\et\big|&=\left|\Im\left[(\psk)^*\left(\eps\nabla\psk-i\et\psk\right)\right]\right|
            \leq|\psk|\,\left|\eps\nabla\psk-i\et\psk\right|\\
      &\leq\left(|\Phk|+|\psk-\Phk|\right)\left(\eps|\nabla\psk-\nabla\Phk|+\e(t)\,|\psk-\Phk|
            +|\eps\nabla\Phk-i\et\Phk|\right)\,.
\end{split}
\end{equation*}
Then by \eqref{eq.scale_Phk}, \eqref{eq.scale_nabla_Phk} and Lemma \ref{lem.pointw_est_psk}
\begin{equation*}
\begin{split}
\big|\bj^{\psk}(\bx,t)&-|\psk(\bx,t)|^2\et\big|\\
      &\leq\left[C\eps^{-\nfourth}\left(1+\frac{|\bx-\at|}{\sqeps}\right)^{k}
            e^{-\nhalf C\left(\frac{|\bx-\at|}{\sqeps}\right)^2}+C\eps^{-\nfourth[1]}\right]\\
      &\hspace{4cm}\left[C\eps^{-\nfourth[1]}+C\eps^{-\nfourth[1]}\left(1+\frac{|\bx-\at|}{\sqeps}\right)^{k+1}
            e^{-\nhalf C\left(\frac{|\bx-\at|}{\sqeps}\right)^2}\right]\\
      &\leq C\left[\eps^{-1}(1+R)^{2k+1}e^{-\nhalf CR^2}+\eps^{-\nhalf}\right]
\end{split}
\end{equation*}
where we have used that $\e(t)$ is continuous and thus bounded on $[0,T]$ and that $(\bx,t)\in\Sigma^\eps_T$ entails $\frac{|\bx-\at|}{\sqeps}=R$. Plugging this into \eqref{eq.|J_U|dsigma}, we see that
\begin{equation*}
\left|J^{\psk}(\bx,t)\cdot\bgu\right|\,d\sigma\leq C\left[(1+R)^{2k+1}e^{-\nhalf CR^2}+\sqeps\right]R^2d\Omega\,.
\end{equation*}
Thus by \eqref{eq.PP(leaves_ball)_leq_FAS}
\begin{equation}
\label{eq.PP(leaves_ball)_leq_gammahalf}
\begin{split}
\PP\left(\mathcal{M}^T_{R\sqeps}\right)
            \leq\int\limits_0^Tdt\int\limits_0^{2\pi}d\phi\int\limits_0^{\pi}d\theta&\sin(\theta)
            CR^2\left[(1+R)^{2k+1}e^{-\nhalf CR^2}+\sqeps\right]\\
      &\leq2\pi T C\left[R^2(1+R)^{2k+1}e^{-\nhalf CR^2}+R^2\sqeps\right]<\nhalf[\gamma]
\end{split}
\end{equation}
for $R$ big and $\eps$ small enough.

Together \eqref{eq.PP(bad_beginning)_leq_gammahalf} and \eqref{eq.PP(leaves_ball)_leq_gammahalf} give the desired result:
\begin{equation*}
\PP\left(\Geps[R]\right)=1-\PP\left((\Geps[R])^c\right)>1-\gamma
\end{equation*}
for all $R$ big and all $\eps$ small enough.


\subsection{Proof of Lemma \ref{lem.pointw_est_psk}}
\label{subsec.proof-lem-ptws-est}
In view of \eqref{eq.def_Ueps} and \eqref{eq.def_Utild}   we have that for all  $t\in[0,T]$
\begin{equation}
\label{eq.cook}
\psk(\bx,t)-\Phk(\bx,t)=\left[\Ueps-\Utild[t,0]\right]\Phk(\bx,0)
      =-\frac{i}{\eps}\int\limits_0^t\Ueps[t-s]V_3(\bx,\at[(s)])\Phk(\bx,s)d\,s\,,
\end{equation}
where $V_3=\Heps-\Htildeps[]$ is the third order remainder term of the the potential's Taylor expansion about $a$.
A priori, equality in \eqref{eq.cook} holds in the sense of $L^2$-functions, i.e.\;for almost every $\bx\in\Rn$, only. In the course of our proof (Lemma \ref{lem.Linfty-est-UepsVmPhk} below) we shall however see that $U^\eps V_3\Phk$ is continuously differentiable with respect to $\bx$ and that $U^\eps V_3\Phk$ and $\nabla U^\eps V_3\Phk$ are bounded for all $s,t\in[0,T]$ and $\bx\in\Rn$. So by dominated convergence also $\psk-\Phk$ (and thus $\psk$) is continuously differentiable with
\begin{equation}
\label{eq.nabla-cook}
\begin{split}
\nabla\psk(\bx,t)-\nabla\Phk(\bx,t)&=-\frac{i}{\eps}\nabla\int\limits_0^t\Ueps[t-s]V_3(\bx,\at[(s)])\Phk(\bx,s)d\,s\\
      &=-\frac{i}{\eps}\int\limits_0^t\nabla\Ueps[t-s]V_3(\bx,\at[(s)])\Phk(\bx,s)d\,s\,.
\end{split}
\end{equation}
Moreover, by continuity \eqref{eq.cook} and \eqref{eq.nabla-cook} hold in fact pointwise for \emph{all} $\bx\in\Rn$.

Our control on $(\nabla)U^\eps V_3\Phk$ is given in
\begin{lem}
\label{lem.Linfty-est-UepsVmPhk}
Let $V\in G_V,\;\bk\in\Nn$ and $T>0$. For $m\in\N$ let
$$
\Vmxt[]{m}:=V(\bx)-\sum_{|\alpha|=0}^{m-1}\frac{1}{\alpha!}\left(\Dalpha V\right)(\ba)(\bx-\ba)^{\alpha}
$$
denote the $m$th remainder term of the Taylor expansion of $V$ about $\ba$.\\
Then $\Ueps[t-s]V_m(\cdot,\at[(s)])\Phk(\cdot,s)$ is continuously differentiable for all $s,\,t\in[0,T]$ and there exists some $C<\infty$ such that
\begin{gather}
\label{eq.Linfty-est-UepsVmPhk}
\max\limits_{s,t\in[0,T]}\left\|\Ueps[t-s]V_m(\cdot\,,\at[(s)])\Phk(\cdot\,,s)\right\|_\infty
        \leq C\eps^{\nhalf[m]-\nfourth}\\
\intertext{and}
\label{eq.Linfty-est-nabla-UepsVmPhk}
\max\limits_{s,t\in[0,T]}\left\|\,|\nabla\Ueps[t-s]V_m(\cdot\,,\at[(s)])\Phk(\cdot\,,s)|\,\right\|_\infty
        \leq C\eps^{\nhalf[m]-\nfourth[7]}\,.
\end{gather}
\end{lem}
Then, plugging \eqref{eq.Linfty-est-UepsVmPhk} and \eqref{eq.Linfty-est-nabla-UepsVmPhk} into \eqref{eq.cook} and \eqref{eq.nabla-cook} immediately yields Lemma \ref{lem.pointw_est_psk}, i.e.
\begin{gather*}
\max_{t\in[0\,,T]}\left\|\psk(\cdot\,,t)-\Phk(\cdot\,,t)\right\|_{\infty}\leq\frac{T}{\eps}
      \max\limits_{s,t\in[0,T]}\left\|\Ueps[t-s]V_3(\cdot\,,\at[(s)])\Phk(\cdot\,,s)\right\|_\infty
      \leq CT\eps^{-\nfourth[1]}\\
\intertext{and}
\max_{t\in[0\,,T]}\left\|\,\left|\nabla\psk(\cdot\,,t)-\nabla\Phk(\cdot\,,t)\right|\,\right\|_{\infty}
      \leq\frac{T}{\eps}
      \max\limits_{s,t\in[0,T]}\left\|\,|\nabla\Ueps[t-s]V_3(\cdot\,,\at[(s)])\Phk(\cdot\,,s)|\,\right\|_\infty
      \leq CT\eps^{-\nfourth[5]}\,.
\end{gather*}

\begin{pro}[of Lemma \ref{lem.Linfty-est-UepsVmPhk}]
First we fix some notation. Let $\geps[\bx]:=\Ueps[t-s]V_m(\bx,\at[(s)]\Phk(\bx,s)$ and $\gepstild[\bx]:=e^{-\frac{i}{\eps}\skpr{\et}{\xa}}\geps[\bx]$.

We shall use an instance of Gagliardo-Nirenberg's inequality \cite{gagliardo:59,nirenberg:59}: For every $n\in\N$ and $l>\nhalf[n]$ there is some $C<\infty$ such that for every $f\in W^{l,2}(\Rn[n]) =\{f\in L^2(\Rn[n])\mid\max\limits_{|\alpha|\leq l}\|D^\alpha f\|_2<\infty\}$
\begin{equation}
\label{eq.gag-niren}
\|f\|_\infty\leq C\Big(\max\limits_{|\alpha|=l}\|D^\alpha f\|_2\Big)^{\frac{n}{2l}}\,\|f\|^{1-\frac{n}{2l}}_2\,.
\end{equation}
Moreover, $f\in C^r(\Rn[n])$ for all $0\leq r<l-\nhalf[n]$.

Applying \eqref{eq.gag-niren} with $n=3$ and $l=2$ to $g^\eps_{m,k}$ gives
\begin{equation}
\label{eq.gag-niren-spec}
\|\geps\|_\infty
      \leq C\left(\max\limits_{|\alpha|=2}\|D^\alpha \geps\|_2\right)^{\frac{3}{4}}\,\|\geps\|^{\frac{1}{4}}_2
\end{equation}
for all $t,s\in[0,T]$. Thus we get \eqref{eq.Linfty-est-UepsVmPhk} if $\|D^\alpha g^\eps_{m,\bk}\|_2=\eps^{-|\alpha|}\|\bp^\alpha U^\eps V_m\Phk\|_2\sim\eps^{\nhalf[m-|\alpha|]}$ for all $\alpha\in\Nn$ with $|\alpha|\in\{0,2\}$.

Unfortunately the latter is generally  false. This is due to the fact that in order to have $\bp\Phk=-i\eps\nabla\Phk\sim\be\Phk$ (i.e.\ part two of \eqref{eq.higher_mom_Phk}), the $\Phk$'s must possess an appropriate, fast varying phase factor. Indeed, since, roughly, $V_m(\bx,\ba)\sim (\bx-\ba)^m$ and $\frac{\bx-\ba}{\sqeps}$ acts on $\Phk$ as a combination of lowering and raising operators (cf.\ \eqref{eq.expansion-(x-a/sqeps)Phk}),
$$
V_m\Phk\sim\eps^{\nhalf[m]}\sum_{|\bk'-\bk|\leq m}\Phk[\bk']
$$
and thus even $\|D^\alpha V_m\Phk\|_2 \sim\eps^{\nhalf[m]-|\alpha|}\|\bp^\alpha\Phk[\bk']\|_2 \sim\eps^{\nhalf[m]-|\alpha|}\|\be^\alpha\Phk[\bk']\|_2 \sim\eps^{\nhalf[m]-|\alpha|}$ is of order $\eps^{-\nhalf[|\alpha|]}$ worse than what we need.

To account for this, we substract the problematical phase, that is we use Gagliardo-Nirenberg not on $g^\eps_{m,\bk}=U^\eps V_m\Phk$ itself but on $\tilde{g}^\eps_{m,\bk}=e^{-\frac{i}{\eps}\skpr{\e}{\xa[]}}U^\eps V_3\Phk$. Then instead of \eqref{eq.gag-niren-spec} we get
\begin{gather*}
\|\geps\|_\infty=\|\gepstild\|_\infty
      \leq C\max\limits_{|\alpha|=2}\|D^\alpha \gepstild\|_2^{\frac{3}{4}}\,\|\gepstild\|^{\frac{1}{4}}_2\\
\intertext{with the higher order terms}
\|\Dalpha\gepstild\|_2 =\eps^{-|\alpha|}\|\left(\bp-\et\right)^\alpha\geps\|_2\,.
\end{gather*}
So \eqref{eq.Linfty-est-UepsVmPhk} holds if $\|(\bp-\e)^\alpha g^\eps_{m,\bk}\|_2\sim\eps^{\nhalf[m+|\alpha|]}$ for all $\alpha\in\Nn$ with $|\alpha|\in\{0,2\}$. Analogously one sees that \eqref{eq.Linfty-est-nabla-UepsVmPhk} holds if $\|(\bp-\e)^\alpha g^\eps_{m,\bk}\|_2\sim\eps^{\nhalf[m+|\alpha|]}$ for all $\alpha\in\Nn$ with $|\alpha|\leq 3$. However, that these estimates for $(\bp-\e)^\alpha g^\eps_{m,\bk}$ hold true is the content of Lemma \ref{lem.L2-est-UepsVmPhk} below.
\end{pro}

\begin{rem}
\label{rem.what-if-canon-sobol}
Instead of the Gagliardo-Nirenberg inequality \eqref{eq.gag-niren} one could also use canonical Sobolev inequalities. However, then one gets results that are not of optimal order in $\eps$, that is instead of Lemma \ref{lem.Linfty-est-UepsVmPhk} one only gets
\begin{gather*}
\left\|U^\eps V_m\Phk\right\|_\infty\leq C\Big[\sum\limits_{|\alpha|=0}^{2}
      \eps^{-|\alpha|}\left\|(\bp-\et)^\alpha U^\eps V_m\Phk\right\|^2_2\Big]^{\nhalf}
      \leq \tilde{C}\eps^{\nhalf[m]-1}\\
\intertext{and}
\|\,|\nabla U^\eps V_m\Phk|\,\|_\infty\leq C\Big[\sum\limits_{|\alpha|=0}^{3}
    \eps^{-|\alpha|}\left\|(\bp-\et)^\alpha U^\eps V_m\Phk\right\|^2_2\Big]^{\nhalf}
      \leq \tilde{C}\eps^{\nhalf[m]-2}\,.
\end{gather*}
Note that also these weaker results suffice to get convergence to classical behavior in the sense of our theorem -- but with a lower rate of convergence. More precisely, instead of \eqref{eq.P(Q_good)>1-gamma} one gets
\begin{equation*}
\PP(\{\bx\in\Rn\mid\,\max\limits_{t\in[0,T]}|\Qp-\at|\leq R\eps^{\nfourth[1]}\})>1-\gamma\,.
\end{equation*}
\end{rem}

\begin{lem}
\label{lem.L2-est-UepsVmPhk}
Let $V\in G_V$. For every $T>0,\,m\in\N$ and $\bk\in\Nn$ there exists some $C<\infty$ such that
\begin{equation}
\label{eq.L2-est-UepsVmPhk}
\max\limits_{s,t\in[0,T]}\left\|\left(\bp-\et\right)^\alpha\Ueps[t-s]V_m(\cdot\,,\at[(s)])\Phk(\cdot\,,s)\right\|_2
        \leq C\eps^{\nhalf[m+|\alpha|]}
\end{equation}
for all multi-indices $0\leq|\alpha|\leq3$.
\end{lem}

\begin{rem}
Since $\psk(\bx,t)=\Ueps\Phk(\bx,0)$, by replacing $\Vmxt[(s)]{m}\Phk(\bx,s)$ with $\Phk(\bx,s)$ in the proof of Lemma \ref{lem.L2-est-UepsVmPhk} and setting $s=0$ one can easily show that also
\[
\max\limits_{t\in[0,T]}\left\|\left(\bp-\et\right)^\alpha\psk(\cdot\,,t)\right\|_2\leq C\eps^{\nhalf[|\alpha|]}
\]
for some $C<\infty$ and all $0\leq|\alpha|\leq3$. So we have, for example, that regarding momentum not only the $\Phk(\bx,t)$'s but also the $\psk(\bx,t)$'s standard deviation is of order $\sqeps$. Since the momentum operator $\bp$ is unbounded this is not a consequence of Hagedorn's results $\|\psk-\Phk\|_2\sim\sqeps$ and $\left\|\left(\bp-\be\right)^\alpha\Phk\right\|_2\sim\eps^{\nhalf[|\alpha|]}$ \cite{hagedorn:85,hagedorn:98}.
\end{rem}

\begin{pro}[of Lemma \ref{lem.L2-est-UepsVmPhk}]
We expand the notation of Lemma \ref{lem.Linfty-est-UepsVmPhk}: For any $l\leq m\in\N$ let
\begin{gather*}
\Vmxt[]{l,m}:=\Vmxt[]{l}-\Vmxt[]{m+1}
      =\sum_{|\alpha|=l}^{m}\frac{1}{\alpha!}\left(\Dalpha V\right)(\ba)(\bx-\ba)^{\alpha}\\
\intertext{and}
\begin{aligned}
\feps[\bx]&:=\Vmxt[(s)]{m}\Phk(\bx,s)&\text{resp.}\qquad&f^\eps_{(m,l),\bk}(\bx,s):=\Vmxt[(s)]{m,l}\Phk(\bx,s)\,,\\
\geps[\bx]&=\Ueps[t-s]\feps[\bx]&\text{resp.}\qquad&g^\eps_{(m,l),\bk}(\bx,t,s):=\Ueps[t-s]f^\eps_{(m,l),k}(\bx,s)\,.
\end{aligned}
\end{gather*}

In the following we set $\|\cdot\|=\|\cdot\|_2$. We shall first prove the weaker result ($|\alpha|\leq 3$)
\begin{equation}
\label{eq.L2-est-(p-et)^alpha-geps-weak}
\max\limits_{s,t\in[0,T]}\left\|\left(\bp-\et\right)^\alpha\geps\right\|\leq C\eps^{\nhalf[m]}
\end{equation}
and then use a bootstrapping argument to arrive at \eqref{eq.L2-est-UepsVmPhk}.

Since $\et$ is bounded on $[0,T]$, instead of \eqref{eq.L2-est-(p-et)^alpha-geps-weak} it suffices to prove that
\begin{equation}
\label{eq.L2-est-p^alpha-geps}
\max\limits_{s,t\in[0,T]}\left\|\bp^\alpha\geps\right\|\leq C\eps^{\nhalf[m]}
\end{equation}
for some $C<\infty$ and all $|\alpha|\leq3$. For that we first get rid of the (unitary) time evolution $U^\eps$, i.e.\;we shall express $\|\bp^\alpha g^\eps_{m,\bk}\|$ in terms of $\|f^\eps_{m,\bk}\|,\,\|\Heps f^\eps_{m,\bk}\|$ and $\|(\Heps)^2 f^\eps_{m,\bk}\|$. We then mimic the proof of (2.38) in \cite{hagedorn:98} to find estimates for the latter.

Since $U^\eps$ is unitary
\begin{equation}
\label{eq.L2(geps)=L2(feps)}
\|\geps\|=\|\feps\|\,.
\end{equation}
Since $\bp=-i\eps\nabla$ is self-adjoint, by Schwarz's inequality and \eqref{eq.L2(geps)=L2(feps)}
\begin{align}
\begin{split}
\label{eq.L2-D1-geps}
\max_{|\alpha|=1}\|\bp^\alpha\geps\|&=\max_{j}\skpr{\geps}{p^2_j\geps}^{\nhalf}\\
      &\hspace{2.5cm}\leq\left(\|\feps\|\,\|p^2\geps\|\right)^{\nhalf}\,,
\end{split}\\
\notag
\max_{|\alpha|=2}\|\bp^\alpha\geps\|&\leq\|p^2\geps\|\\
\intertext{and}
\notag
\max_{|\alpha|=3}\|\bp^\alpha\geps\|&\leq\left(\|p^2\geps\|\|p^4\geps\|\right)^{\nhalf}\,.
\end{align}
Thus we get \eqref{eq.L2-est-p^alpha-geps} if we can show that $\|f^\eps_{m,\bk}\|,\,\|p^2g^\eps_{m,\bk}\|$ and $\|p^4g^\eps_{m,\bk}\|$ are of order $\eps^{\nhalf[m]}$. Write $p^2=2(\Heps-V)$. Since $[\Heps,U^\eps]=0$ and $V$ is bounded by $C_V$ (cf.\;Definition \ref{def.(V)^m}),
\begin{equation}
\label{eq.L2-p^2geps}
\begin{split}
\|p^2\geps\|&=2\|(\Heps-V)\geps\|\\
        &\leq2\left[\|\Heps\Ueps[t-s]\feps\|+\|V\|_\infty\|\geps\|\right]\\
    &\leq2\left[\|\Heps\feps\|+C_V\|\feps\|\right]\,.
\end{split}
\end{equation}
In the same way
\begin{equation*}
\begin{split}
\|p^4\geps\|&=4\|(\Heps-V)^2\geps\|\\
   &\leq4\bigg[\|(\Heps)^2\feps\|+2\|V\|_\infty\|\Heps\feps\|+\|V\|_\infty^2\|\feps\|\\
        &\qquad\qquad+\|[\Heps,V]\geps\|\bigg]\\
   &\leq4\bigg[\|(\Heps)^2\feps\|+2C_V\|\Heps\feps\|+C_V^2\|\feps\|\\
        &\qquad\qquad+\eps\|\skpr{\nabla V}{\bp}\geps\|
                +\nhalf[\eps^2]\|\triangle V\|_\infty\|\feps\|\bigg]
\end{split}
\end{equation*}
Since $V\in G_V$ implies that also $\nabla V$ and $\triangle V$ are bounded by $C_V$, this yields
\begin{equation*}
\begin{split}
\|p^4&\geps\|\\
      &\stackrel{\eqref{eq.L2-D1-geps}}{\leq}4\bigg[\|(\Heps)^2\feps\|+2C_V\|\Heps\feps\|
                  +C_V(C_V+\nhalf[\eps^2])\|\feps\|\\
            &\hspace{6.7cm}+3\eps C_V\left(\|\feps\|\,\|p^2\geps\|\right)^{\nhalf}\bigg]\\
      &\stackrel{\eqref{eq.L2-p^2geps}}{\leq}4\bigg[\|(\Heps)^2\feps\|+2C_V\|\Heps\feps\|
                  +C_V(C_V+\nhalf[\eps^2])\|\feps\|\\
            &\hspace{4cm}+3\sqrt{2}\eps C_V\|\feps\|^{\nhalf}\big(\|\Heps\feps\|+C_V\|\feps\|\big)^{\nhalf}\bigg]\,.
\end{split}
\end{equation*}

Thus we get \eqref{eq.L2-est-p^alpha-geps} if we can show that $\|f^\eps_{m,\bk}\|,\,\|\Heps f^\eps_{m,\bk}\|$ and $\|(\Heps)^2 f^\eps_{m,\bk}\|$ are of order $\eps^{\nhalf[m]}$. We mimic the proof of (2.38) in \cite{hagedorn:98} and introduce the following splitting ($R>0$):
\begin{equation*}
\|\feps\|^2=\hspace{-0.5cm}\int\limits_{|\xa[(s)]|\leq R}\hspace{-0.5cm}\big|V_m(\bx,\at[(s)])\Phk(\bx,s)\big|^2d^3x
      \;+\hspace{-0.5cm}\int\limits_{|\xa[(s)]|> R}\hspace{-0.5cm}\big|V_m(\bx,\at[(s)])\Phk(\bx,s)\big|^2d^3x
      =:\ro{I}\,+\,\ro{II}\,.
\end{equation*}
Remember that $V_m$ is the remainder
\begin{equation*}
V_m(\bx,\ba)=V(\bx)-\sum\limits_{|\alpha|=0}^{m-1}\frac{1}{\alpha!}(\Dalpha V)(\ba)(\xa[])^\alpha
      =\sum\limits_{|\alpha|=m}\frac{1}{\alpha!}(\Dalpha V)\left(\bxi(\bx,\ba)\right)(\xa[])^\alpha
\end{equation*}
where $\bxi(\bx,\ba)=\ba+\lambda(\xa[])$ for some $\lambda\in(0,1)$.
Thus
\begin{equation*}
\left|V_m(\bx,\at[(s)])\right|
      \leq  \begin{cases}
            \|V\|_\infty+\max\limits_{|\alpha|\leq m-1}\left|(\Dalpha V)(\at[(s)])\right|
                  \sum\limits_{l=0}^{m-1}\sum\limits_{|\alpha|=l}\frac{l!}{\alpha!}\,\frac{|\xa[(s)]|^l}{l!}
                  &\text{ for }\bx\in\Rn\,,\\
            \max\limits_{\xi\leq a(s)+R}\left|(\Dalpha V)(\bxi)\right|\,3^m\frac{|\xa[(s)]|^m}{m!}
                  &\text{ for }|\xa[(s)]|\leq R
            \end{cases}
\end{equation*}
and, since $\at[(s)]$ is continuous in $s$ and $V\in\Gv$ is bounded and $C^\infty$, there exists some $C<\infty$ such that
\begin{equation*}
\left|V_m(\bx,\at[(s)])\right|
      \leq  \begin{cases}
            C\sum\limits_{l=0}^{m-1}\frac{(3|\xa[(s)]|)^l}{l!}\leq C e^{3|\xa[(s)]|} & \text{ for }\bx\in\Rn\,,\\
            C |\xa[(s)]|^m & \text{ for }|\xa[(s)]|\leq R
            \end{cases}
\end{equation*}
for all $s\in[0,T]$. Substituting $\by:=\frac{\xa[(s)]}{\sqeps}$, with this and \eqref{eq.scale_Phk} we get for all $\eps$ small enough
\begin{gather*}
\ro{I}\leq C\int\limits_{0}^{\frac{R}{\sqeps}}\eps^{m}y^{2m}(1+y)^{2k}\,e^{-Cy^2}dy\,\leq C\eps^m\\
\intertext{and}
\ro{II}\leq C\int\limits_{\frac{R}{\sqeps}}^{\infty}(1+y)^{2k}\,e^{-y(Cy-6\sqeps)}dy\,\leq C e^{-\frac{C}{\sqeps}}\,.
\end{gather*}
So
\begin{equation}
\label{eq.feps_order_eps^mhalf}
\max\limits_{s\in[0,T]}\|\feps\|=\mathcal{O}(\eps^{\nhalf[m]})\,.
\end{equation}

To estimate $\|\Heps f^{\eps}_{m,\bk}\|$ write
\begin{equation*}
\Heps f^\eps_{m,\bk}=V_m\Heps\Phk+[\Heps,V_m]\Phk\,.
\end{equation*}
With $E_{cl}=\nhalf\e^2+\Vat[]$ and $[\Heps,V_m]=-i\eps\skpr{\nabla V_m}{\bp}-\nhalf[\eps^2](\triangle V_m)$ this gives
\begin{align*}
\Heps f^\eps_{m,\bk}&=E_{cl}V_m\Phk+V_m(\Heps-E_{cl})\Phk-i\eps\skpr{\nabla V_m}{\bp}\,\Phk
            -\nhalf[\eps^2](\triangle V_m)\Phk\\
      &=E_{cl}V_m\Phk+\nhalf V_m(p^2-\e^2)\Phk+V_m(V(\bx)-V(\ba))\Phk\\
            &\hspace{3cm}-i\eps\skpr{\nabla V_m}{\be}\,\Phk-i\eps\skpr{\nabla V_m}{\bp-\be}\,\Phk
            -\nhalf[\eps^2](\triangle V_m)\Phk\\
      &=E_{cl}V_m\Phk+\nhalf V_m(\bp-\be)^2\Phk+V_m\skpr{\be}{\bp-\be}\Phk+V_mV_1\Phk\\
            &\hspace{3cm}-i\eps\skpr{\nabla V_m}{\be}\,\Phk-i\eps\skpr{\nabla V_m}{\bp-\be}\,\Phk
            -\nhalf[\eps^2](\triangle V_m)\Phk\,.
\end{align*}
Now, by \eqref{eq.expansion-(p-e/sqeps)Phk} we see that $(\bp-\be)\Phk$ is $\sqeps$ times a (vector of) linear combination(s) of $\Phk[\bk']$'s with $|\bk-\bk'|=1$ and $(\bp-\be)^2\Phk$ is $\eps$ times a linear combination of $\Phk[\bk']$'s with $|\bk-\bk'|\in\{0,2\}$. Thus $\Heps f^\eps_{m,\bk}$ is a sum of terms of the form
\[
C(\be)\tilde{f}^\eps_{m,\bk'}:=C(\be)\tilde{V}^\eps_m\Phk[\bk']
\]
where $C(\be)$ is either a constant or some function of $\be$, $|\bk-\bk'|\leq 2$ and $\tilde{V}^\eps_m$  is a wild card for $V_m,\,\eps V_m,$ $\sqeps V_m,\,V_mV_1,\, \eps(\partial_j V_m),\,\eps^{\nhalf[3]}(\partial_jV_m)$ or $\eps^2(\partial^2_j V_m)$ ($j=1,2,3$). Note that
\begin{equation*}
\Dalpha V_m=(\Dalpha V)_{m-|\alpha|}\,,
\end{equation*}
so $\tilde{V}^\eps_m$ is either $V_mV_1$ or of the form $\eps^{\nhalf[l]}\tilde{V}_{m-r}$ where the new $\eps$-independent ``potential'' $\tilde{V}$ is a wild card for $V,\,\partial_jV$ or $\partial^2_jV$ and $l,r\in\N$ are such that $l-r\geq0$. Now, since $V\in G_V$ implies $\tilde{V}\in C^\infty$ and $\|\tilde{V}\|_\infty\leq\max\limits_{|\alpha|\leq2}\|\Dalpha V\|_\infty\leq C_V$, not only the proof of $\|V_mV_1\Phk[\bk']\|=\mathcal{O}(\eps^{\nhalf[m+1]})$ but also that of $\|\tilde{V}_{m-r}\Phk[\bk']\|=\mathcal{O}(\eps^{\nhalf[m-r]})$ is completely analogous to that of \eqref{eq.feps_order_eps^mhalf}. Therefore, $\|\tilde{f}^\eps_{m,\bk'}\|$ is either of order $\eps^{\nhalf[m+1]}$ (if $\tilde{V}^\eps_m=V_mV_1$) or of order $\eps^{\nhalf[m+l-r]}\leq\eps^{\nhalf[m]}$ (if $\tilde{V}^\eps_m=\eps^{\nhalf[l]}\tilde{V}_{m-r}$), that is we get
\begin{equation}
\label{eq.L2_Heps_feps_O(eps^mhalf)}
\max\limits_{s\in[0,T]}\|\Heps\feps\|\leq\sum\max\limits_{s\in[0,T]}|C(\be)|\,\|\tilde{f}^\eps_{m,\bk'}(\cdot,s)\|
      =\mathcal{O}(\eps^{\nhalf[m]})\,.
\end{equation}

Finally, $\|(\Heps)^2f^\eps_{m,\bk}\|=\mathcal{O}(\eps^{\nhalf[m]})$ clearly follows if we can show that, for each of the above $\tilde{f}^\eps_{m,\bk'},$ $\|\Heps\tilde{f}^\eps_{m\bk'}\|$ is (at least) of order $\eps^{\nhalf[m]}$. The proof of the latter, however, is completely analogous to that of \eqref{eq.L2_Heps_feps_O(eps^mhalf)}. Just note that this time we get up to fourth order derivatives of $V$ as new ``potentials'' $\tilde{V}$, which is why in the definition of $G_V$ we required that $\|\Dalpha V\|_\infty\leq C_V$ for $|\alpha|\leq4$.

So we have shown that \eqref{eq.L2-est-p^alpha-geps} and thus also \eqref{eq.L2-est-(p-et)^alpha-geps-weak} holds. To get \eqref{eq.L2-est-UepsVmPhk} we split off the lowest order term of $V_m$, $V_m=V_{m,m}+V_{m+1}$ (cf.\ notation at the beginning of this proof). Then by \eqref{eq.L2-est-(p-et)^alpha-geps-weak}
\begin{equation}
\label{eq.split-gepsm=geps{mm}+geps(m+1)}
\begin{split}
\|\left(\bp-\et\right)^\alpha\geps\|&\leq\|\left(\bp-\et\right)^\alpha g^\eps_{(m,m),k}(\cdot,t,s)\|
            +\|\left(\bp-\et\right)^\alpha g^\eps_{m+1,\bk}(\cdot,t,s)\|\\
      &\leq\|\left(\bp-\et\right)^\alpha g^\eps_{(m,m),\bk}(\cdot,t,s)\|+C\eps^{\nhalf[m+1]}\,.
\end{split}
\end{equation}
To estimate $\left(\bp-\be\right)^\alpha g^\eps_{(m,m),\bk}$ note that
\[
\begin{split}
g^\eps_{(m,m),\bk}(\bx,t,s)&=\Ueps[t-s]V_{m,m}(\bx,\at[(s)]\Phk(\bx,s)\\
      &=\eps^{\nhalf[m]}\Ueps[t-s]\sum\limits_{|\beta|=m}\frac{1}{\beta!}
            \left(\Dalpha[\beta] V\right)\left(a(s)\right)\left(\frac{\xa[(s)]}{\sqeps}\right)^\beta\Phk(x,s)
\end{split}
\]
and that $\left(\frac{\xa[]}{\sqeps}\right)^\beta\Phk$ is a finite sum of $\Phk[\bk']$s with $|\bk-\bk'|\leq m$ and coefficients that are independent of $\eps$ and bounded on $[0,T]$ (\eqref{eq.expansion-(x-a/sqeps)Phk} and $\At[s]$ continuous in $s$). Since also $\left(\Dalpha[\beta] V\right)\left(\ba(s)\right)$ is bounded on $[0,T]$ ($V\in C^\infty(\Rn)$ and $\ba(s)$ continuous in $s$) it thus suffices to estimate
\[
\eps^{\nhalf[m]}\left(\bp-\et\right)^\alpha\Ueps[t-s]\Phk[\bk'](\bx,s)
\]
for $|\bk-\bk'|\leq m$. Like in \eqref{eq.cook}
\begin{align*}
\Ueps[t-s]\Phk[\bk'](\bx,s)&=\Phk[\bk'](\bx,t)-\frac{i}{\eps}
            \int\limits_s^t\Ueps[t-\tau]V_3\left(\bx,\ba(\tau)\right)\Phk[\bk'](\bx,\tau)\,d\tau\\
      &=\Phk[\bk'](\bx,t)-\frac{i}{\eps}
            \int\limits_s^tg^\eps_{3,\bk'}(\bx,t,\tau)\,d\tau\,.
\end{align*}
Since by \eqref{eq.L2-est-(p-et)^alpha-geps-weak} $\|\left(\bp-\et\right)^\alpha g^\eps_{3,\bk'}(\cdot,t,\tau)\|<C\eps^{\nhalf[3]}$, changing the order of differentiation ($\bp=-i\eps\nabla$) and integration in
\[
\|\left(\bp-\et\right)^\alpha\int\limits_s^t g^\eps_{3,\bk'}(\cdot,t,\tau)\,d\tau\|
      =\|\int\limits_s^t\left(\bp-\et\right)^\alpha g^\eps_{3,\bk'}(\cdot,t,\tau)\,d\tau\|
\]
is justified by dominated convergence and we thus get (for any $s,t\in[0,T]$)
\begin{align*}
\eps^{\nhalf[m]}\|\left(\bp-\et\right)^\alpha&\Ueps[t-s]\Phk[\bk'](\cdot,s)\|\\
      &\leq\eps^{\nhalf[m]}\|\left(\bp-\et\right)^\alpha\Phk[\bk'](\cdot,t)\|
            +\eps^{\nhalf[m]-1}\int\limits_s^t\|\left(\bp-\et\right)^\alpha g^\eps_{3,\bk'}(\cdot,t,\tau)\|\,d\tau\\
      &\leq\eps^{\nhalf[m+|\alpha|]}\left\|\left(\frac{\bp-\et}{\sqeps}\right)^\alpha\Phk[\bk'](\cdot,t)\right\|
            +\eps^{\nhalf[m+1]}CT\,.
\end{align*}
By \eqref{eq.higher_mom_Phk} this yields
\begin{gather*}
\eps^{\nhalf[m]}\|\left(\bp-\et\right)^\alpha\Ueps[t-s]\Phk[\bk'](\cdot,s)\|
      \leq C\left(\eps^{\nhalf[m+|\alpha|]}+\eps^{\nhalf[m+1]}\right)\\
\intertext{and thus also}
\|\left(\bp-\et\right)^\alpha g^\eps_{(m,m),\bk}(\cdot,t,s)\|
      \leq C\left(\eps^{\nhalf[m+|\alpha|]}+\eps^{\nhalf[m+1]}\right)\,.
\end{gather*}
Putting this into \eqref{eq.split-gepsm=geps{mm}+geps(m+1)} we see that we can sharpen \eqref{eq.L2-est-(p-et)^alpha-geps-weak} to
\[
\max\limits_{s,t\in[0,T]}\left\|\left(\bp-\et\right)^\alpha\geps\right\|
      \leq C\left(\eps^{\nhalf[m+|\alpha|]}+\eps^{\nhalf[m+1]}\right)\leq C\eps^{\nhalf[m+1]}\,.
\]
Iterating this bootstrapping argument several times we finally arrive at
\[
\max\limits_{s,t\in[0,T]}\left\|\left(\bp-\et\right)^\alpha\geps\right\|
      \leq C\left(\eps^{\nhalf[m+|\alpha|]}+\eps^{\nhalf[m+|\alpha|]}\right)\,,
\]
i.e.\;at \eqref{eq.L2-est-UepsVmPhk}.

\end{pro}



\ifx\undefined\bysame
\newcommand{\bysame}{\leavevmode\hbox to3em{\hrulefill}\,}
\fi

\end{document}